\documentclass[10pt,prb,superscriptaddress,showpacs,twocolumn]{revtex4}
\usepackage{amssymb,amsmath}
\usepackage{graphicx}
\usepackage{color}
\usepackage{natbib}
\usepackage{hyperref}
\usepackage{floatrow}
\usepackage[T2A,OT1]{fontenc}
\usepackage[utf8]{inputenc}

\begin{document}

\newcommand\cyrtext[1]{{\fontencoding{T2A}\selectfont #1}}  
\newcommand{\df}[2]{\frac{\partial #1}{\partial #2}}             
\newcommand{\ds}[2]{\frac{{\partial}^2 #1}{\partial {#2}^2}}     
\newcommand{\tens}[1]{\bar{\bar{#1}}}   

\title{Nonlocality in uniaxially polarizable media}
\author{Maxim~A.~Gorlach}
\email{Maxim.Gorlach.blr@gmail.com}
\affiliation{ITMO University, St. Petersburg 197101, Russia}
\affiliation{Belarusian State University, Minsk~220030,~Belarus}
\author{Pavel~A.~Belov}
\affiliation{ITMO University, St. Petersburg 197101, Russia}

\begin{abstract}
We reveal extraordinary electromagnetic properties for a general class of uniaxially polarizable media.
Depending on parameters, such metamaterials may have wide range of nontrivial shapes of isofrequency contours including lemniscate, diamond and multiply connected curves with connectivity number reaching five. The possibility of the dispersion engineering paves a way to more flexible manipulation of electromagnetic waves. Employing first-principle considerations we prove that uniaxially polarizable media should be described in terms of the nonlocal permittivity tensor which by no means can be reduced to local permittivity and permeability even in the long-wavelength limit. We introduce an alternative set of local material parameters including quadrupole susceptibility capable to capture all of the second-order spatial dispersion effects.
\end{abstract}

\maketitle


    \begin{figure}[b]
    \includegraphics[width=0.7\linewidth]{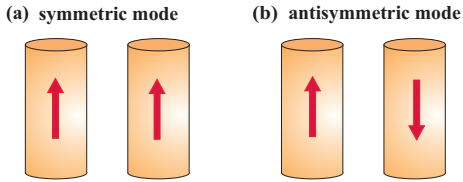}
    \caption{(Color online) The sketch of (a) symmetric and (b) antisymmetric modes in coupled rod pair  structure. Arrows show the direction of the currents.}
    \label{ris:CW}
    \end{figure}

There exists a class of uniaxially polarizable media (UPM) that includes metamaterials based on coupled rod pairs, dual wires, etc.~[\onlinecite{Shalaev2005,Zhou2006,Podolskiy2005}] exhibiting such interesting properties as artificial magnetism and backward waves~[\onlinecite{Soukoulis2007, ShalaevNP2007}]. The basic idea of coupled rod pair metamaterial design is rather simple. The incident electromagnetic wave excites currents in the coupled rods.  
The excitation of the symmetric mode when currents have the same magnitude and direction is usually attributed to the electric dipole resonance of the unit cell (Fig.~\ref{ris:CW}a). The excitation of the antisymmetric mode when currents have the same magnitude but the opposite directions is attributed to the magnetic dipole resonance of the unit cell~(Fig.~\ref{ris:CW}b). Thus, one may conclude that the structure possesses resonant artificial dielectric and magnetic properties due to the presence of the symmetric and antisymmetric modes of the unit cell, respectively. Effective medium model (EMM) for coupled rod pair metamaterials and similar structures in terms of local permittivity $\tens{\varepsilon}^{\rm{loc}}$ and permeability $\tens{\mu}^{\rm{loc}}$ tensors was introduced in the theoretical works Refs.~[\onlinecite{Panina,Pets2008, Pshenay2011}], verified by numerical simulations~[\onlinecite{Huang2010}] and by experiments~[\onlinecite{Dolling2005,Guven,Shalaev2005,ZhouPRB,Yuan2007}] based on the Nicolson-Ross-Weir extraction procedure~[\onlinecite{Nicolson,Weir}].

In this Communication, we demonstrate that the exact description of the entire class of uniaxially polarizable media is incompatible with the local effective medium model, introduce an alternative set of local material parameters and reveal the unusual dispersion properties of UPM. The isofrequency contours that are  lemniscate-like, diamond-shaped and multiply connected (with connectivity number equal to five) appear if $\partial^2\varepsilon_{zz}/\partial k_z^2>0$, $\varepsilon_{zz}(\omega,0)\geq 0$ and $0<\partial^2\varepsilon_{zz}/\partial k_y^2<2 c^2/\omega^2$ or $\partial^2\varepsilon_{zz}/\partial k_y^2<0$, respectively.

The nonlocal permittivity tensor~[\onlinecite{landau8,Silv2007}] of UPM consisting of reciprocal elements with inversion symmetry can be reduced to~\footnote{Once $P_x\equiv P_y\equiv 0$, uniaxially polarizable medium is characterized by $\varepsilon_{xy}=\varepsilon_{xz}=\varepsilon_{yx}=\varepsilon_{yz}=0$ and $\varepsilon_{xx}=\varepsilon_{yy}=1$. We assume here that $\bar{\bar{\varepsilon}}(\omega,-\vec{k})=\bar{\bar{\varepsilon}}(\omega,\vec{k})$, i.e. structure under the study possesses inversion symmetry. Under this assumption and employing the symmetry of the kinetic coefficients~[\onlinecite{landau5,landau8}], it can be shown that $\varepsilon_{ik}(\omega,\vec{k})=\varepsilon_{ki}(\omega,\vec{k})$, i.e. the effective nonlocal permittivity tensor is symmetric. Therefore, $\varepsilon_{zx}=\varepsilon_{zy}=0$. As a result, the only nontrivial component of the nonlocal permittivity tensor is the $zz$ component.}
\begin{equation}\label{GeneralPermittivity}
\bar{\bar{\varepsilon}}(\omega,\vec{k})=
\begin{pmatrix}
1 & 0 & 0\\
0 & 1 & 0\\
0 & 0 & \varepsilon_{zz}(\omega,\vec{k})
\end{pmatrix}\:,
\end{equation}
where the $z$ axis is chosen in such way that $\vec{P}=P_z\,\mathbf{z}$. Within the frame of EMM, the effective nonlocal permittivity tensor can be written in terms of $\tens{\varepsilon}^{\rm{loc}}$ and $\tens{\mu}^{\rm{loc}}$ as follows~[\onlinecite{Silv2007,Vinogradov}]:
\begin{equation}\label{EffMedPermittivity}
\bar{\bar{\varepsilon}}_{\rm{eff}}(\omega,\vec{k})=\bar{\bar{\varepsilon}}^{\rm{loc}}(\omega)+\frac{1}{q^2}\,\vec{k}\times\left((\bar{\bar{\mu}}^{\rm{loc}}(\omega))^{-1}-\bar{\bar{I}}\right)\times\vec{k}\:,
\end{equation}
where $q=\omega/c$, $c$ is the speed of light, and the bianisotropic terms are excluded since they vanish due to the structure inversion symmetry. Comparing Eq.~\eqref{GeneralPermittivity} and Eq.~\eqref{EffMedPermittivity} it is easy to show that uniaxially polarizable medium is described by EMM in the only case when $\tens{\mu}^{\rm{loc}}=\tens{I}$.~\footnote{Indeed, comparing Eqs.~\eqref{GeneralPermittivity}, \eqref{EffMedPermittivity} we obtain
$
\vec{k}\times\bar{\bar{A}}\times\vec{k}=\Delta(\vec{k})\,{\mathbf z}{\mathbf z}$
where ${\bf z}$ denotes a unit vector along the $z$ axis, $\bar{\bar{A}}=\left((\bar{\bar{\mu}}^{\rm{loc}}(\omega))^{-1}-\bar{\bar{I}}\right)/q^2$, and the term $\Delta(\vec{k})\equiv \varepsilon_{zz}(\omega,\vec{k})-\varepsilon_{zz}(\omega,0)$ is the spatial dispersion corrections to $\varepsilon_{zz}$. This matrix equation yields eight scalar equations with the zero right hand side. Taking into account that the wave vector components $k_x$, $k_y$ and $k_z$ are independent, one obtains a homogeneous system of nine linear equations with the nonzero determinant. Thus, we infer that the only possible solution for the matrix $\bar{\bar{A}}$ is a zero matrix and, as a consequence, $\bar{\bar{\mu}}^{\rm{loc}}=\tens{I}$. } 

    \begin{figure}[hb!]
    \includegraphics[width=0.65\linewidth]{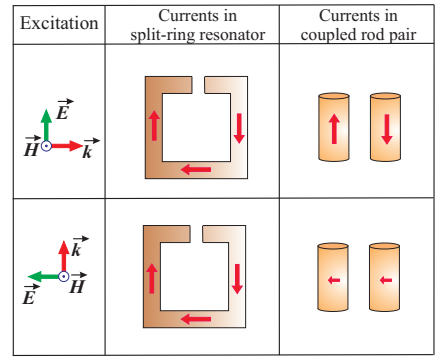}
    \caption{(Color online) Illustration of the difference in magnetic response of SRR and coupled rod pair.}
    \label{ris:Explanation}
    \end{figure}
    
In the other words, local permeability different from unity cannot be attributed to uniaxially po\/la\/ri\/zab\/le media. The qualitative explanation of the obtained result is as follows. We consider the excitation of the structure (Fig.~\ref{ris:Explanation}) by the waves of two types that have the same direction of magnetic field but different directions of electric field and wave vector. In the case of split-ring resonator (SRR),  the response to these two excitations is determined mainly by the direction of magnetic field and therefore is almost the same. On the other hand, coupled rod pair responds differently to these excitations as the alignment of the electric field vector with respect to the pair axis is crucial. This simple reasoning suggests that the magnetic response of coupled rod pair structure cannot be described properly by the local permeability tensor.

Thus, it is important to introduce an alternative set of local parameters giving the UPM consistent description. From now on we assume that for the UPM description  it is sufficient to consider only spatial dispersion of the second order and that higher order nonlocal effects are negligible. Under this approximation we may expand $\varepsilon_{zz}(\omega,\vec{k})$ in Taylor series with respect to $\vec{k}$. Due to the presumed inversion symmetry the first-order terms of the expansion vanish and it acquires the form:
\begin{equation}\label{EpsExpansion}
\varepsilon_{zz}(\omega,\vec{k})=\varepsilon^{\rm{loc}}(\omega)+\sum_{i,j}\,\chi_{ij}(\omega)\,k_i k_j\:,
\end{equation}
where $\varepsilon^{\rm{loc}}=\varepsilon_{zz}(\omega,0)$. The quadrupole susceptibility matrix $\chi_{ij}(\omega)=\frac{1}{2}\,\frac{\partial^2 \varepsilon_{zz}}{\partial k_i\,\partial k_j}$ depends only on the frequency and is symmetric with respect to the indices $i$ and $j$. Thus, the quadrupole susceptibility $\tens{\chi}$ yields an alternative set of local material parameters, the number of independent components being equal to six due to the matrix $\tens{\chi}$ symmetry (the same as in the symmetric permeability tensor). Using Eq.~\eqref{EpsExpansion}, it is easy to obtain the relation for the macroscopic fields in space-time domain applicable to the general UPM:
\begin{equation}\label{ConsEq}
\begin{split}
& D_{x}=E_{x}\:, \mspace{10mu} D_y=E_y\:,\\
& D_{z}=\varepsilon^{\rm{loc}}\,E_z-\sum\limits_{i,j}\chi_{ij}\,\partial_i\partial_j E_z\:.
\end{split}
\end{equation}

    \begin{figure}[b!]
     \caption{(Color online) The typical isofrequency contours for uniaxially polarizable medium with $\chi_{zz}<0$. The plane $k_x=0$ is examined.}
    \includegraphics[width=0.9\linewidth]{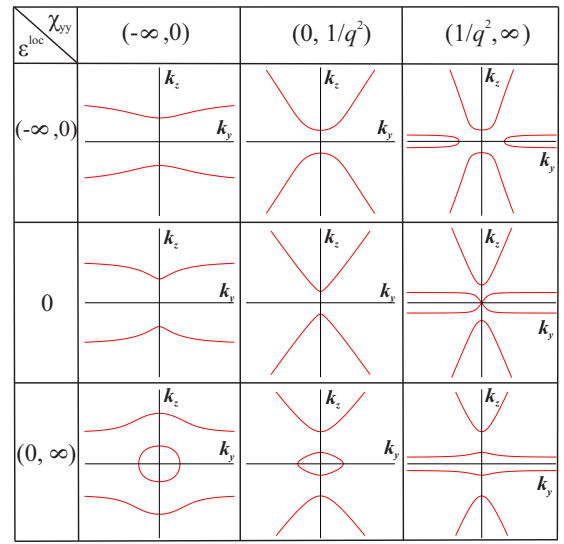}
    \label{ris:IsofrM}
    \end{figure}
    
The symmetries of the structure reduce the number of the matrix $\tens{\chi}$ independent components~[\onlinecite{Jones}]. The nonzero components of quadrupole susceptibility for structures of different symmetries are summarized in the Table~\ref{tab:Chi}.
For the sake of simplicity we consider the UPM with the symmetry $C_2$, i.e. the structure invariant under rotation at the angle $\pi$  around the anisotropy axis $z$. For example, this is the case for the coupled rod pair structure. The quadrupole susceptibility is given by the matrix

\begin{table}[ht!]
\begin{tabular}{|c|p{0.5\linewidth}|p{0.25\linewidth}|}
\hline
Symmetry & Nonzero components of $\tens{\chi}$ & Independent components \\
\hline
No & $\chi_{xx}$, $\chi_{yy}$, $\chi_{zz}$, $\chi_{xy}=\chi_{yx}$, $\chi_{xz}=\chi_{zx}$, $\chi_{zy}=\chi_{yz}$ & 6\\
\hline
$C_2$ & $\chi_{xx}$, $\chi_{yy}$, $\chi_{zz}$, $\chi_{xy}=\chi_{yx}$ & 4\\
\hline
$C_3$ or $C_4$ & $\chi_{xx}=\chi_{yy}$, $\chi_{zz}$ & 2\\
\hline
\end{tabular}
\caption{The nonzero components of quadrupole susceptibility for the structures of different symmetries}
\label{tab:Chi}
\end{table}

\begin{equation}\label{ChiTens}
\tens{\chi}
=\begin{pmatrix}
\chi_{xx} & \chi_{xy} & 0 \\
\chi_{xy} & \chi_{yy} & 0 \\
0 & 0 & \chi_{zz}
\end{pmatrix}\:,
\end{equation}
where the components depend on the geometry of the particular material and on the frequency. Now we consider the dispersion equation for TM waves in UPM
\begin{equation}\label{DispEquation}
k_x^2+k_y^2=(q^2-k_z^2)\,\varepsilon_{zz}(\omega,\vec{k})
\end{equation}
with $\varepsilon_{zz}(\omega,\vec{k})$ given by Eqs.~\eqref{EpsExpansion}, \eqref{ChiTens}. To illustrate physical effects possible in UPM, we study the propagation in the plane $Oyz$ when $k_x=0$, and in this case Eq.~\eqref{DispEquation} reduces to
\begin{equation}\label{DispEquation1}
k_y^2\,\left[1-(q^2-k_z^2)\,\chi_{yy}\right]=(q^2-k_z^2)\,\left(\varepsilon^{\rm{loc}}+\chi_{zz}k_z^2\right)\:.
\end{equation}

    \begin{figure}[t]
    \caption{(Color online) The typical isofrequency contours for uniaxially polarizable medium with $\chi_{zz}>0$. The plane $k_x=0$ is examined.}
    \includegraphics[width=0.9\linewidth]{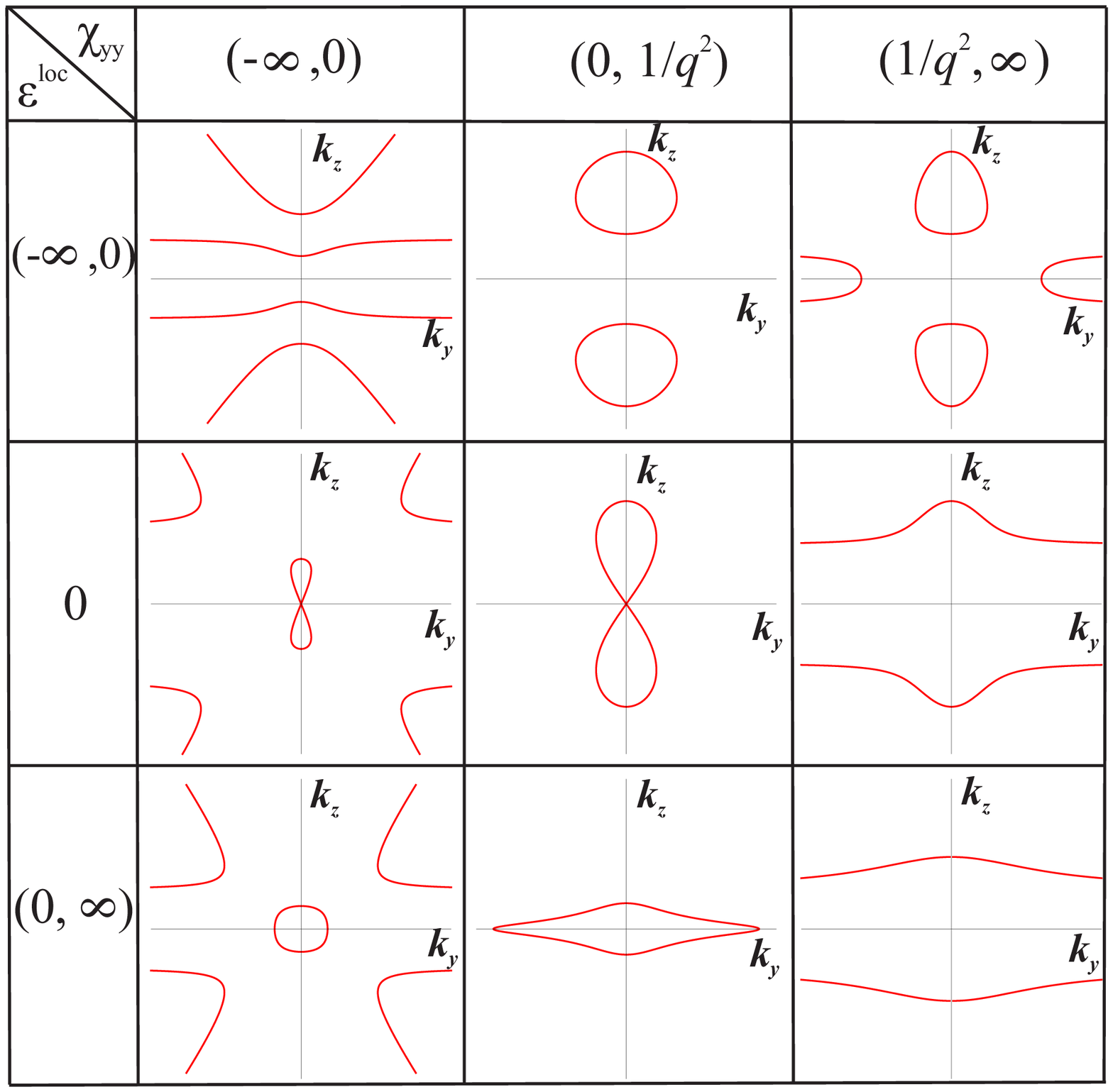}
    \label{ris:IsofrP}
    \end{figure}

    \begin{figure}[ht]
    \caption{(Color online) The geometry of the extraction experiment for uniaxially polarizable medium. TM polarization of the incident wave is explored. Coupled rods are parallel to the interface.}
    \includegraphics[width=0.8\linewidth]{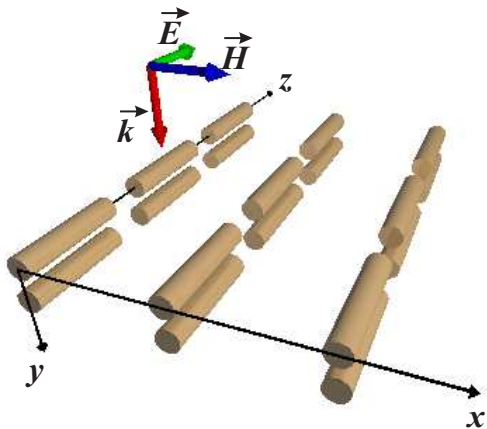}
    \label{ris:Boundary}
    \end{figure}

The equation of the similar structure is valid for the plane of propagation $k_x=k_y=k_{\bot}/\sqrt{2}$. In such case, $k_y$ should be replaced by $k_{\bot}$, and $\chi_{yy}$ by $1/2\,\left(\chi_{xx}+\chi_{yy}+2\chi_{xy}\right)$ in Eq.~\eqref{DispEquation1}.

In conventional local media with anisotropic uniaxial permittivity and permeability tensors, one can obtain elliptic and hyperbolic isofrequency contours, i.e. the second-order curves~[\onlinecite{Smith2003}]. Equation~\eqref{DispEquation1} yields that the isofrequency contours in UPM are the fourth order curves with the connectivity number more than two. This paves a way to more flexible electromagnetic waves manipulation using the nonlocal effects. Solving Eq.~\eqref{DispEquation1} we can visualise the dispersion properties of UPM. The system of isofrequency contours for various values of $\varepsilon^{\rm{loc}}$, $\chi_{zz}$ and $\chi_{yy}$ is plotted in Figs.~\ref{ris:IsofrM} and \ref{ris:IsofrP}. As one can see, UPM's exhibit interesting and unusual dispersion properties. Among the curves presented in Figs.~\ref{ris:IsofrM}, \ref{ris:IsofrP} there are both isofrequency contours that have already been observed for particular systems and the isofrequency contours that to our knowledge have not been observed so far for any particular structure. Namely, the isofrequency contour Fig.~\ref{ris:IsofrM} for $\varepsilon^{\rm{loc}}>0$ and $0<\chi_{yy}<1/q^2$ is the same as one in the low-$\varepsilon$ mixed dispersion regime in the structure composed of the uniaxial electric scatterers~[\onlinecite{Gorlach2014}]. Curves Fig.~\ref{ris:IsofrM} for $\chi_{yy}<0$ have a similarity with the isofrequency contours obtained for the structure composed of dielectric rods~[\onlinecite{Silv2006}]. A lemniscate-like isofrequency contour arises in the regime $\chi_{zz}>0$ (Fig.~\ref{ris:IsofrP}), $\varepsilon^{\rm{loc}}\approx 0$ and $0<\chi_{yy}<1/q^2$. The structures with such dispersion properties can be useful for obtaining superradiance or for phase matching (if nonlinear elements are present). A diamond-shaped isofrequency contour appears if $\chi_{zz}>0$, $\varepsilon(\omega,0)>0$ and $0<\chi_{yy}<1/q^2$; such dispersion properties are useful in Purcell factor engineering~[\onlinecite{ChebykinP}].

It is also important that the dispersion equation Eq.~\eqref{DispEquation1} is linear with respect to $k_y^2$ and biquadratic with respect to $k_z$. Therefore, if a surface of the structure is normal to $y$ axis (Fig~\ref{ris:Boundary}), a single transmitted beam will be observed and the spatial dispersion effects are hidden to some extent. However, if the surface of the structure is perpendicular to the anisotropy axis $z$, two transmitted TM modes will be excited which serves as a direct ma\/ni\/fes\/ta\/tion of the nonlocal effects. 

To determine the type of the dispersion regime for the particular UPM one has to infer the quadrupole susceptibility matrix $\tens{\chi}$ of the structure. This can be fulfilled employing the technique similar to Nicolson-Ross-Weir method~[\onlinecite{Nicolson,Weir}]. In the case $k_z=0$ (Fig.~\ref{ris:Boundary}) Eq.~\eqref{DispEquation1} reduces to the following one:
\begin{equation}\label{DispEquation2}
k_y^2=q^2\,\frac{\varepsilon^{\rm{loc}}}{1-q^2\,\chi_{yy}}\:.
\end{equation}
Comparing this result to the standard formula $k_y^2=q^2\,\varepsilon^{\rm{loc}}\,\mu^{\rm{loc}}$ we conclude that it is possible to introduce local ``permeability'' for this particular direction of propagation:
\begin{equation}\label{Muloc}
\mu^{\rm{loc}}=\frac{1}{1-q^2\,\chi_{yy}}\:.
\end{equation}
If the function on the right-hand side of Eq.~\eqref{DispEquation2} is decreasing with the frequency, the wave described by Eq.~\eqref{DispEquation2} is backward. It is this backward wave that was observed in coupled rod pair metamaterials~[\onlinecite{ShalaevNP2007}]. Moreover, assuming the sectional continuity of the fields at the UPM surface and integrating Maxwell equations over a pillbox, it can be easily demonstrated that the boundary conditions for the TM wave incident at the UPM surface from vacuum are equivalent to those for the boundary between magnetic \eqref{Muloc} and vacuum in the special case of normal incidence ($k_x=k_z=0$ in the geometry Fig.~\ref{ris:Boundary}):
\begin{equation}\label{BoundaryCond}
\begin{split}
& E_{1z}=E_{2z}\;,\\
& B_{1x}=\left(\mu^{\rm{loc}}\right)^{-1}\,B_{2x}\:.
\end{split}
\end{equation}
that corresponds to the boundary conditions for the conventional magnetic with the local parameters $\varepsilon^{\rm{loc}}$ and $\mu^{\rm{loc}}$, and the reflection coefficients determined in the case of normal incidence will coincide with those for the magnetic Eq.~\eqref{Muloc}. Thus, $\varepsilon^{\rm{loc}}$ and $\mu^{\rm{loc}}$ can be extracted by the standard Nicolson-Ross-Weir technique~[\onlinecite{Guven,Shalaev2005,Yuan2007}], and $\chi_{yy}$ can be calculated via $\mu^{\rm{loc}}$ as $\chi_{yy}=\left(1-(\mu^{\rm{loc}})^{-1}\right)/q^2$. The analogous approach is valid for $\chi_{xx}$ extraction; for this purpose the plane of the pairs shold be parallel to the boundary of the structure. However, in the case of oblique incidence with $k_z\not =0$ EMM is no longer valid. From the measured reflection at oblique incidence it is possible to infer $\chi_{zz}$. Investigation of the reflection of the $TM$ wave incident in the plane different from $Oyz$ and $Oxy$ (namely, the plane $k_x=k_y$) will allow one to extract $\chi_{xy}$ component of the structure quadrupole susceptibility. The detailed discussion of UPM quadrupole susceptibility extraction will be published elsewhere.

The reasoning similar to the performed above reveals that for the structures that have the effective nonlocal permittivity tensor of the form
\begin{equation}\label{TwoDirEps}
\tens{\varepsilon}(\omega,\vec{k})=
\begin{pmatrix}
\varepsilon_{xx} & \varepsilon_{xy} & 0 \\
\varepsilon_{yx} & \varepsilon_{yy} & 0 \\
0 & 0 & \varepsilon_{zz}
\end{pmatrix}
\end{equation}
the only possible nontrivial component of the magnetic permeability tensor is $\mu_{zz}$ component, i.e. $\tens{\mu}=\mathbf{x}\mathbf{x}+\mathbf{y}\mathbf{y}+\mu_{zz}\,\mathbf{z}\mathbf{z}$. It is interesting to notice that the fishnet structure~[\onlinecite{Soukoulis2007}] made from reciprocal material comprises the class of materials described by Eq.~\eqref{TwoDirEps}. Indeed, $\varepsilon_{zx}=\varepsilon_{zy}=0$ and $\varepsilon_{zz}=1$ for the fishnet structure that can be polarized in the plane $Oxy$. Making use of the symmetry of the effective permittivity tensor, we obtain also that $\varepsilon_{xz}=\varepsilon_{yz}=0$. Therefore, quadrupole susceptibilities should be also used for the description of the structures Eq.~\eqref{TwoDirEps}.

In this work, we reveal the intriguing dispersion properties of uniaxially polarizable media, namely,  nontrivial shapes of isofrequency contours including lemniscate, diamond and multiply connected curves with connectivity number reaching five. All these unusual properties are beyond effective medium model. Problems arising in the effective medium approach were illustrated in a number of works~[\onlinecite{Agranovich2006,Costa,Simovski2010,Menzel2010}]. In the present Communication, we demonstrate incompleteness of local effective medium model for the entire class of uniaxially polarizable media proving from the first principles that these structures do not possess magnetic properties ($\tens{\mu}^{\rm{loc}}=\tens{I}$) though they do exhibit second-order spatial dispersion effects. We also propose an alternative set of  local material parameters that are capable to capture all of the second-order nonlocal effects. This set of parameters allows one to demonstrate how the backward waves arise in uniaxially polarizable media due to the spatial dispersion effects and provide a correct interpretation to the experiments extracting material parameters of uniaxially polarizable media.

We acknowledge fruitful discussions with C.R.~Simovski and A.N.~Poddubny. We are also grateful to Mehedi Hasan for his interest in this work. The present work was supported by the Government of the Russian Federation (Grant No. 074-U01), grant of the President of the Russian Federation No. \cyrtext{МД}-6805.2013.2, Russian Foundation for Basic Research (grant No.~15-02-08957 A), ``Dynasty'' foundation, the Ministry of Education and Science of the Russian Federation (projects 14.584.21.0009 10, GOSZADANIE 2014/190, Zadanie no. 3.561.2014/K)  and the Ministry of Education of the Republic of Belarus (grant No.~625/02).

\bibliography{Bibliography}
\end{document}